\documentclass[amsmath,amssymb,prb]{revtex4}

\usepackage{graphics}

\newcommand{\tr}{{\mathrm{tr}}}
\begin{document}
\title{Many-body theory of degenerate systems: A simple example}

\author{Christian Brouder}
\affiliation{%
Laboratoire de Min\'eralogie Cristallographie, CNRS UMR 7590,
Universit\'es Paris 6 et 7, IPGP, case 115, 4 place Jussieu,
75252 Paris cedex 05, France.
}%

\date{\today}

\begin{abstract}
The hierarchy of Green functions for (quasi)degenerate
systems, presented in cond-mat/0308058, is calculated 
in detail for the case of a system with closed shells
plus a single electron in a two-fold degenerate level.
The complete hierarchy is derived explicitly, in terms
of Green functions and of Feynman diagrams.
\end{abstract}

%
\maketitle
\newcommand{\un}{{\mathbf{1}}}

\newcommand{\ee}{{\mathrm{e}}}
\newcommand{\dd}{{\mathrm{d}}}
\newcommand{\Id}{{\mathrm{Id}}}
\newcommand{\inter}{{\mathrm{int}}}
\newcommand{\degres}{{\mathrm{deg}}}
\newcommand{\barT}{{T^*}}
\newcommand{\calA}{{\cal{A}}}
\newcommand{\calD}{{\cal{D}}}
\newcommand{\calL}{{\cal{L}}_I}
\newcommand{\calH}{{\cal{H}}}
\newcommand{\action}{{\cal{A}^\inter}}
\newcommand{\counit}{{\varepsilon}}
\newcommand{\deltaj}{{\delta j}}
\newcommand{\barun}{{\bar{1}}}
\newcommand{\barunp}{{\bar{1}'}}
\newcommand{\barn}{{\bar{n}}}
\newcommand{\barnp}{{\bar{n}'}}
\newcommand{\barde}{{\bar{2}}}
\newcommand{\bardep}{{\bar{2}'}}
\newcommand{\bartr}{{\bar{3}}}
\newcommand{\bartrp}{{\bar{3}'}}
\newcommand{\barx}{{\bar{x}}}
\newcommand{\barxp}{{\bar{x}'}}
\newcommand{\baru}{{\bar{u}}}
\newcommand{\barv}{{\bar{v}}}
\newcommand{\bara}{{\bar{\alpha}}}
\newcommand{\bareta}{{\bar{\eta}}}
\newcommand{\barpsi}{{\bar{\psi}}}

\newcommand{\bfn}{{\mathbf{n}}}
\newcommand{\bfk}{{\mathbf{k}}}
\newcommand{\bfr}{{\mathbf{r}}}
\newcommand{\bfun}{{\mathbf{1}}}
\newcommand{\bfX}{{\mathbf{X}}}
\newcommand{\bflambda}{{\boldsymbol{\lambda}}}
\newcommand{\bfrho}{{\boldsymbol{\rho}}}
\newcommand{\bfmu}{{\boldsymbol{\mu}}}

\renewcommand{\i}[1]{{}_{\scriptscriptstyle(#1)}}
\newcommand{\iu}[1]{{}_{\scriptscriptstyle(\underline #1)}}
\newcommand{\Deltau}{\underline{\Delta}}

\newcommand{\bc}{{\bar{c}}}
\newcommand{\bd}{{\bar{d}}}

\section{Introduction}
In reference \cite{BrouderKB}, hereafter referred to as I, 
a many-body theory of degenerate or quasidegenerate systems
was presented, using quantum group methods.
In particular, a hierarchy of Green function was 
derived and formulated in terms of reduced coproducts.
In this paper, the general formula is written explicitly
in terms of Green functions and Feynman diagrams.
The case we consider is that of a system with closed
shells and a two-fold degenerate level occupied by
a single electron.  This does not look like a very
interesting example, but we shall see that the
hierarchy involves $n$-body Green functions with
$n$ up to 5. In the general case of a $M$-fold degenerate
level, the Green functions involved go up to 
$n=3M-1$. In practice, such complicated Green functions
are not usable, and the hierarchy is stopped after
the first terms. The hierarchy we give here contains
the first terms of the general hierarchy.

Quantum group
concepts are not (yet) familiar in solid-state physics,
so the pace will be slow and the calculations will be
presented in full detail.

\section{Calculation of $W^1_\rho$}
The simplest case is one electron ($N=1$)
in a twofold degenerate level ($M=2$).
The two degenerate levels are denoted by $1$ and $2$.
The density matrix is
\begin{eqnarray*}
\bfrho &=&
\left( \begin{array}{cc}
    \rho_{11} & \rho_{12} \\
   \rho_{21} & \rho_{22} \end{array}\right),
\end{eqnarray*}
with $\tr(\bfrho)= \rho_{11}+ \rho_{22}=1$.

The moment generating function $\rho(\bara,\alpha)$ is
defined in equation (\ref{rho(baraalpha)}) of I. In our
case, $N=1$ and
$\rho(\bara,\alpha)=\rho_0(\bara,\alpha)+\rho_1(\bara,\alpha)$.
According to equation (\ref{rhoN}) of I (with  $N=1$, $i_1=1$ or $2$,
$j_1=1$ or $2$)
\begin{eqnarray*}
\rho_1(\bara,\alpha) &=& \rho_{11} \bara_1\alpha_1 +
              \rho_{12} \bara_1\alpha_2 +
              \rho_{21} \bara_2\alpha_1 +
              \rho_{22} \bara_2\alpha_2.
\end{eqnarray*}
Using now equation (\ref{rhok}) of I
with $k=1$ we obtain
\begin{eqnarray*}
\rho_0(\bara,\alpha) &=& \tr(\bfrho) = \rho_{11}+\rho_{22}=1.
\end{eqnarray*}
According to equation (\ref{rhoc(baraalpha)}) of
I, the cumulant generating function
$\rho^c(\bara,\alpha)$ is obtained from
\begin{eqnarray*}
\log\big(\rho(\bara,\alpha)\big) &=&
\log\big(1+\rho_1(\bara,\alpha)\big)
=
\rho_1(\bara,\alpha) +\rho^c(\bara,\alpha),
\end{eqnarray*}
where 
\begin{eqnarray*}
\rho^c(\bara,\alpha) &=& \rho^c_2(\bara,\alpha)
=-\frac{\rho_1(\bara,\alpha)^2}{2 }
=-\det(\bfrho) \bara_1\alpha_1\bara_2\alpha_2
=
-(\rho_{11}\rho_{22}-\rho_{12}\rho_{21})
\bara_1\alpha_1\bara_2\alpha_2.
\end{eqnarray*}
According to equation (\ref{Wrho0fin}) of I, this
enables us to calculate $W^0_\rho$. But we saw in
section \ref{W0W1} of I that $W^1_\rho=W^0_\rho$, thus
\begin{eqnarray}
W_\rho^1 &=& 
-i\int \bareta(x) G^0_\rho(x,y)\eta(y)\dd x\dd y
+\rho^c_2(\bara,\alpha),
\label{Wrho1}
\end{eqnarray}
with
\begin{eqnarray*}
G^0_\rho(x,y) &=& G^0_0(x,y) 
+i \Big(\sum_{m=1}^C  u_m(x) \baru_m(y)+
   \sum_{k=1}^2\sum_{l=1}^2 \rho_{lk} u_k(x) \baru_l(y)
\Big) M,
\end{eqnarray*}
where
\begin{eqnarray*}
M &=& \left( \begin{array}{cc}
    1 & -1 \\
   -1 & 1 \end{array}\right).
\end{eqnarray*}
The expression for the $2\times 2$ matrix $G^0_0(x,y)$
is given in equation (\ref{freeGreen}) of I.

For later convenience, we give the identities
\begin{eqnarray*}
\frac{\delta^2 \bara_1\alpha_1\bara_2\alpha_2}
{\delta \eta_{\epsilon'}(y)
\delta \bareta_{\epsilon}(x)}
&=&\epsilon\epsilon'
(u_1(x)\bara_2-u_2(x)\bara_1)
(\baru_1(y)\alpha_2-\baru_2(y)\alpha_1).
\end{eqnarray*}
\begin{eqnarray}
\frac{\delta^4 \rho^c_2(\bara,\alpha)}
{\delta \bareta_{\epsilon_1}(x_1)
\delta \bareta_{\epsilon_2}(x_2)
\delta \eta_{\epsilon'_1}(y_1)
\delta \eta_{\epsilon'_2}(y_2)}
&=&-\epsilon_1\epsilon_2\epsilon'_1\epsilon'_2 
r(x_1,x_2,y_1,y_2),
\label{delta4}
\end{eqnarray}
where
\begin{eqnarray*}
 r(x_1,x_2,y_1,y_2) &=& 
-\det(\bfrho)
(u_1(x_1) u_2(x_2)-u_2(x_1) u_1(x_2))
(\baru_1(y_1)\baru_2(y_2)-\baru_2(y_1)\baru_1(y_2)).
\end{eqnarray*}

\section{Hierarchy}
For one electron in a two-fold degenerate state,
equation (\ref{reduit}) of I stops at $n=3$ because
$M=2$ and we have
\begin{eqnarray*}
\frac{\delta Z_\rho}{\delta\beta} &=&
\frac{\delta W^1_\rho}{\delta\beta} Z_\rho
-i \sum (-1)^{|D\i{1'}|} \big(D\i{1'} 
\frac{\delta W^1_\rho}{\delta\beta}\big)
\big(D\i{2'} Z_\rho \big)
- \frac{1}{2}
\sum (-1)^{|D\i{1'}^{2}|} \big(D\i{1'}^{2} 
\frac{\delta W^1_\rho}{\delta\beta}\big)
\big(D\i{2'}^{2} Z_\rho\big)
\\&&
+ \frac{i}{6}
\sum (-1)^{|D\i{1'}^{3}|} \big(D\i{1'}^{3} 
\frac{\delta W^1_\rho}{\delta\beta}\big)
\big(D\i{2'}^{3} Z_\rho\big).
\end{eqnarray*}
To obtain the Green function, we put
$\beta=\bareta_\epsilon(x)$, we make a 
functional derivative with respect to 
$\eta_{\epsilon'}(y)$ and we set all fermion sources
to zero (in particular, this implies $Z_\rho=1$).
We use also the fact that a functional derivative
of $Z_\rho$ or $W^1_\rho$ with respect to an odd
number of fermion sources is zero if the sources
are set to zero. 
This enables us to calculate
the signs $(-1)^{|D\i{1'}^{n}|}$.
\begin{eqnarray}
\frac{\delta^2 Z_\rho}{\delta\eta_{\epsilon'}(y) \delta\bareta_\epsilon(x)} &=&
\frac{\delta^2 W^1_\rho}{\delta\eta_{\epsilon'}(y) 
        \delta\bareta_\epsilon(x)}
-i \sum 
\big(D\i{1'} \frac{\delta^2 W^1_\rho}{\delta\eta_{\epsilon'}(y) 
             \delta\bareta_\epsilon(x)}\big)
\big(D\i{2'} Z_\rho \big)
-i \sum \big(D\i{1'}
\frac{\delta W^1_\rho}{\delta\bareta_\epsilon(x)}\big)
\big(D\i{2'} \frac{\delta Z_\rho}{\delta\eta_{\epsilon'}(y)} \big)
\nonumber\\&&
- \frac{1}{2}
\sum \big(D\i{1'}^{2}
\frac{\delta^2 W^1_\rho}{\delta\eta_{\epsilon'}(y)
        \delta\bareta_\epsilon(x)}\big)
\big(D\i{2'}^{2} Z_\rho\big)
- \frac{1}{2}
\sum \big(D\i{1'}^{2}
\frac{\delta W^1_\rho}{\delta\bareta_\epsilon(x)}\big)
\big(D\i{2'}^{2} \frac{\delta Z_\rho}{\delta\eta_{\epsilon'}(y)} \big).
\nonumber\\&&
+ \frac{i}{6}
\sum \big(D\i{1'}^{3}
\frac{\delta W^1_\rho}{\delta\bareta_\epsilon(x)}\big)
\big(D\i{2'}^{3} \frac{\delta Z_\rho}{\delta\eta_{\epsilon'}(y)} \big).
\label{reduit2}
\end{eqnarray}

\subsection{Coproduct}

The non-relativistic interacting Hamiltonian is
\begin{eqnarray}
\int H^\inter(t) \dd t &=& 
\frac{e^2}{2} \int \barpsi(t,\bfr) \barpsi(t,\bfr')
\frac{1}{|\bfr-\bfr'|}\psi(t,\bfr') \psi(t,\bfr)  
\dd t \dd \bfr\dd \bfr'.
\label{Hamil}
\end{eqnarray}
To simplify the notation, we define $x_1=(t,\bfr)$,
$x'_1=(t',\bfr')$ and $v(x_1-x'_1)=\delta(t-t')e^2/|\bfr-\bfr'|$,
so that
\begin{eqnarray*}
\int H^\inter(t) \dd t &=& \frac{1}{2} \int v(x_1-x'_1)
\barpsi(x_1) \barpsi(x'_1)\psi(x'_1) \psi(x_1)
\dd x_1 \dd x'_1.
\end{eqnarray*}
To define the interacting operator $D$ (i.e.
the functional differential form of $\int H^\inter(t) \dd t$),
we make the substitutions
$\psi(x_1)\rightarrow -i\delta/\delta\bareta(x_1)$
and
$\barpsi(x_1)\rightarrow i\delta/\delta\eta(x_1)$
and we use the simplified notation
\begin{eqnarray*}
\delta_{1_\pm}&=&\frac{\delta}{\delta \eta_\pm(x_1)},
\quad
\delta_{\barun_\pm}=\frac{\delta}{\delta \bareta_\pm(x_1)},
\quad
\delta_{1'_\pm}=\frac{\delta}{\delta \eta_\pm(x'_1)},
\quad
\delta_{\barunp_\pm}=\frac{\delta}{\delta \bareta_\pm(x'_1)}.
\end{eqnarray*}
This gives us $D=D_+-D_-$ with
\begin{eqnarray}
D_\pm &=&  \frac{1}{2} \int v(x_1-x'_1)
   \delta_{1_\pm}\delta_{1'_\pm}
\delta_{\barunp_\pm}\delta_{\barun_\pm} \dd x_1 \dd x'_1.
\label{Dpm}
\end{eqnarray}
Thus
\begin{eqnarray*}
D &=& \frac{1}{2} \int \dd x_1\dd x'_1 v(x_1-x'_1) 
\Big(
\delta_{1_+}\delta_{1'_+}\delta_{\barunp_+} \delta_{\barun_+}
- \delta_{1_-}\delta_{1'_-}\delta_{\barunp_-} \delta_{\barun_-}
\Big),
\end{eqnarray*}
To further simplify the notation, we replace $\delta_{1_\pm}$ by
$\delta_1$ and $x_1$ by $1$ and $x'_1$ by $1'$.
The interacting operator becomes
\begin{eqnarray*}
D &=& \frac{1}{2} \int \dd 1\dd 1' v(1-1')
\sum_{\pm} \pm \delta_{1}\delta_{1'}\delta_{\barunp} \delta_{\barun}.
\end{eqnarray*}

To calculate the coproduct of $D$, we start from the
coproduct of the basic functional derivatives
\begin{eqnarray*}
\Delta \delta_{1} &=& \delta_{1}\otimes \un
+ \un\otimes \delta_{1},\quad
\Delta \delta_{1'} = \delta_{1'}\otimes \un
+ \un\otimes \delta_{1'},\quad
\Delta \delta_{\barun} = \delta_{\barun}\otimes \un
+ \un\otimes \delta_{\barun},\quad
\Delta \delta_{\barunp} = \delta_{\barunp}\otimes \un
+ \un\otimes \delta_{\barunp}.
\end{eqnarray*}
Then we use equation (\ref{DeltaDD'}) of I to 
calculate the coproduct of 
$ \delta_{1} \delta_{1'}$
and
$\delta_{\barunp}\delta_{\barun}$:
\begin{eqnarray*}
\Delta \delta_{1} \delta_{1'}
&=&  \delta_{1} \delta_{1'} \otimes \un
  + \delta_{1}\otimes \delta_{1'}
  - \delta_{1'}\otimes \delta_{1}
  + \un \otimes \delta_{1} \delta_{1'},\quad
\Delta \delta_{\barunp}\delta_{\barun}
= \delta_{\barunp}\delta_{\barun} \otimes \un
 + \delta_{\barunp}\otimes\delta_{\barun}
 - \delta_{\barun} \otimes  \delta_{\barunp}
 + \un \otimes \delta_{\barunp}\delta_{\barun}.
\end{eqnarray*}
We use equation (\ref{DeltaDD'}) of I again
to obtain
\begin{eqnarray}
\Delta \delta_{1} \delta_{1'}
       \delta_{\barunp}\delta_{\barun}
&=& \delta_{1} \delta_{1'}
     \delta_{\barunp}\delta_{\barun} \otimes \un
  + \un\otimes \delta_{1} \delta_{1'}
     \delta_{\barunp}\delta_{\barun}
  +  \delta_{1} \delta_{1'}
     \delta_{\barunp}\otimes\delta_{\barun}
  - \delta_{1} \delta_{1'}
      \delta_{\barun} \otimes  \delta_{\barunp}
  + \delta_{1} \delta_{1'} \otimes
      \delta_{\barunp}\delta_{\barun}
  +  \delta_{1} \delta_{\barunp}\delta_{\barun} 
       \otimes \delta_{1'}
\nonumber\\&&
  -  \delta_{1}\delta_{\barunp}\otimes
       \delta_{1'}\delta_{\barun}
  +  \delta_{1} \delta_{\barun} \otimes 
       \delta_{1'} \delta_{\barunp}
  + \delta_{1}\otimes \delta_{1'}
       \delta_{\barunp}\delta_{\barun}
  -  \delta_{1'}\delta_{\barunp}\delta_{\barun} 
       \otimes \delta_{1}
  +  \delta_{1'} \delta_{\barunp} \otimes
       \delta_{1} \delta_{\barun}
  -  \delta_{1'} \delta_{\barun} \otimes
       \delta_{1} \delta_{\barunp}
\nonumber\\&&
  - \delta_{1'}\otimes \delta_{1}
       \delta_{\barunp}\delta_{\barun}
  + \delta_{\barunp}\delta_{\barun} \otimes 
       \delta_{1} \delta_{1'}
 + \delta_{\barunp}\otimes \delta_{1}
         \delta_{1'}\delta_{\barun}
  - \delta_{\barun}\otimes \delta_{1}
       \delta_{1'} \delta_{\barunp}.
\label{coprodD}
\end{eqnarray}
The reduced coproduct with respect to $D$ is obtained by
its definition $\Delta'D=\Delta D-
 D\otimes \un - \un\otimes D$. Therefore,
\begin{eqnarray}
\Delta' D &=& \frac{1}{2} \int \dd 1\dd 1' v(1-1')
\sum_{\pm} \pm  \Big(
    \delta_{1} \delta_{1'}
     \delta_{\barunp}\otimes\delta_{\barun}
  - \delta_{1} \delta_{1'}
      \delta_{\barun} \otimes  \delta_{\barunp}
  + \delta_{1} \delta_{1'} \otimes
      \delta_{\barunp}\delta_{\barun}
  +  \delta_{1} \delta_{\barunp}\delta_{\barun} 
       \otimes \delta_{1'}
\nonumber\\&&
  -  \delta_{1}\delta_{\barunp}\otimes
       \delta_{1'}\delta_{\barun}
  +  \delta_{1} \delta_{\barun} \otimes 
       \delta_{1'} \delta_{\barunp}
  + \delta_{1}\otimes \delta_{1'}
       \delta_{\barunp}\delta_{\barun}
  -  \delta_{1'}\delta_{\barunp}\delta_{\barun} 
       \otimes \delta_{1}
  +  \delta_{1'} \delta_{\barunp} \otimes
       \delta_{1} \delta_{\barun}
  -  \delta_{1'} \delta_{\barun} \otimes
       \delta_{1} \delta_{\barunp}
\nonumber\\&&
  - \delta_{1'}\otimes \delta_{1}
       \delta_{\barunp}\delta_{\barun}
  + \delta_{\barunp}\delta_{\barun} \otimes 
       \delta_{1} \delta_{1'}
 + \delta_{\barunp}\otimes \delta_{1}
         \delta_{1'}\delta_{\barun}
  - \delta_{\barun}\otimes \delta_{1}
       \delta_{1'} \delta_{\barunp}\Big).
\label{coprodpD}
\end{eqnarray}
Notice that, in $\Delta'D=\sum D\i{1'}\otimes D\i{2'}$, each term
$D\i{1'}$ or $D\i{2'}$ contains between one and three functional 
derivatives.

\subsection{Expansion of equation (\ref{reduit2})}
Now we are going to examine each term of equation
(\ref{reduit2}).  We define the $n$-body Green function as
\begin{eqnarray}
G_{\epsilon_1\dots\epsilon_n\epsilon'_1\dots\epsilon'_n}
(x_1,\dots,x_n,y'_1,\dots,y'_n)
&=&
(-i)^n  \frac{\delta^{2n} Z_\rho }
{\delta\bareta_{\epsilon_1}(x_1)\cdots \delta\bareta_{\epsilon_n}(x_n)
\delta\eta_{\epsilon'_1}(y_1)\cdots \delta\eta_{\epsilon'_n}(y_n)}.
\label{nbodyG}
\end{eqnarray}
The right hand side is taken at zero external sources.
To simplify the notation, we write $i$ for $x_i$ 
and $i'$ for $y_i$, we consider that $\epsilon_i$ goes with the
variable $x_i$, $\epsilon'_i$ with $y_i$, so the $\epsilon$ are now 
implicit and we have
\begin{eqnarray*}
G(1,\dots,n,1',\dots,n')
&=&
(-i)^n \,\delta_{\barun}\dots \delta_{\barn}
    \delta_{\barunp}\dots \delta_{\barnp}
    Z_\rho.
\end{eqnarray*}
With this notation, the left hand side of equation  (\ref{reduit2})
is now
\begin{eqnarray*}
\frac{\delta^2 Z_\rho}{\delta\eta_{\epsilon'}(y) \delta\bareta_\epsilon(x)} &=&
-\frac{\delta^2 Z_\rho}{\delta\bareta_\epsilon(x)\delta\eta_{\epsilon'}(y)}=
-i G(x,y).
\end{eqnarray*}
More explicitly 
$G(x,y)=G_{\epsilon\epsilon'}(x,y)$.
In the following we shall calculate in detail all the
terms of the right hand side of equation (\ref{reduit2}).

\subsection{First term of equation (\ref{reduit2})}
The first term of (\ref{reduit2}) is 
very simple. From equation (\ref{Wrho1}), we know that
$W_\rho^1$ is the sum of a term of degree two and a term
of degree four in the external sources.
The first term in the right hand side of equation (\ref{reduit2})
contains only two functional derivatives. Thus, at zero sources,
the second term of $W_\rho^1$ does not contribute.
By definition of functional derivatives, the first term gives us
\begin{eqnarray*}
\frac{\delta^2 W^1_\rho}{\delta\eta_{\epsilon'}(y) 
        \delta\bareta_\epsilon(x)} &=& -i 
G^0_{\rho\epsilon\epsilon'}(x,y).
\end{eqnarray*}

\subsection{Second term of equation (\ref{reduit2})}
The second term of equation (\ref{reduit2}) is
\begin{eqnarray*}
-i \sum 
\big(D\i{1'} \delta_y \delta_\barx W^1_\rho\big)
\big(D\i{2'} Z_\rho \big),
\end{eqnarray*}
where we wrote $\delta_y$ for $\delta/\delta\eta_{\epsilon'}(y)$ and
$\delta_\barx$ for $\delta/\delta\bareta_{\epsilon}(x)$.
According to equations (\ref{coprodpD}),
each term $D\i{1'}$ contains at least one functional derivative.
Therefore, the first term of $W^1_\rho$ does not contribute
because it is of degree 2. The second term of $W^1_\rho$ is of degree
four, and it contains two sources $\eta$ and two sources $\bareta$.
The functional derivatives are evaluated at zero sources, thus
$D\i{1'}$ must be the product of a derivative with respect to a source 
$\eta$ and a derivative with respect to a source $\bareta$.
According to equation (\ref{coprodpD}), there are four terms 
with this property in $\Delta'D$:
\begin{eqnarray*}
\frac{1}{2} \int \dd 1\dd 1' v(1-1')
\sum_{\pm} \pm  \Big(
  -  \delta_{1}\delta_{\barunp}\otimes
       \delta_{1'}\delta_{\barun}
  +  \delta_{1} \delta_{\barun} \otimes 
       \delta_{1'} \delta_{\barunp}
  +  \delta_{1'} \delta_{\barunp} \otimes
       \delta_{1} \delta_{\barun}
  -  \delta_{1'} \delta_{\barun} \otimes
       \delta_{1} \delta_{\barunp}\Big).
\end{eqnarray*}

The second term of equation (\ref{reduit2}) becomes
\begin{eqnarray}
&&
-\frac{i}{2} \int \dd 1 \dd 1' v(1-1') \sum_{\pm}
\pm\Big( 
-\big(\delta_{1}\delta_{\barunp}\delta_{y}
  \delta_{\barx} W^1_\rho\big)
 \big(\delta_{1'}\delta_{\barun} Z_\rho\big)
+\big(\delta_{1}\delta_{\barun}\delta_{y}
  \delta_{\barx} W^1_\rho\big)
 \big(\delta_{1'}\delta_{\barunp} Z_\rho\big)
+\big(\delta_{1'}\delta_{\barunp}\delta_{y}
  \delta_{\barx} W^1_\rho\big)
 \big(\delta_{1}\delta_{\barun} Z_\rho\big)
\nonumber\\&&\hspace*{43mm}
-\big(\delta_{1'}\delta_{\barun}\delta_{y}
  \delta_{\barx} W^1_\rho\big)
 \big(\delta_{1}\delta_{\barunp} Z_\rho\big)\Big).
\label{secondterm}
\end{eqnarray}
Let us now consider the first term of (\ref{secondterm}).
According to (\ref{nbodyG}),
$\delta_{1'}\delta_{\barun} Z_\rho=-iG(1,1')$.
According to (\ref{delta4}),
\begin{eqnarray*}
\delta_{1}\delta_{\barunp}\delta_{y} \delta_{\barx} W^1_\rho
&=& \delta_{\barx}\delta_{\barunp}\delta_{1}\delta_{y} W^1_\rho
= -\epsilon\epsilon' r(x,1',1,y).
\end{eqnarray*}
For the second term of (\ref{secondterm}) we have 
$\delta_{1'}\delta_{\barunp} Z_\rho = -i G(1',1')$
and
\begin{eqnarray*}
\delta_{1}\delta_{\barun}\delta_{y} \delta_{\barx} W^1_\rho
&=& \delta_{\barx}\delta_{\barun}\delta_{1}\delta_{y} W^1_\rho
= -\epsilon\epsilon' r(x,1,1,y).
\end{eqnarray*}
Interchanging the variables $1$ and $1'$, we see that the
third term of (\ref{secondterm}) is equal to the first one, 
and the fourth term is equal to the second one.
Finally, the second term of (\ref{reduit2}) becomes
the sum of
\begin{eqnarray*}
&&
-\epsilon\epsilon' \int \dd 1 \dd 1' v(1-1') r(x,1',1,y)
\sum_{\pm} \pm G(1,1'),
\end{eqnarray*}
and
\begin{eqnarray*}
\epsilon\epsilon' \int \dd 1 \dd 1' v(1-1') r(x,1,1,y)
\sum_{\pm} \pm G(1',1').
\end{eqnarray*}

\subsection{Third term of equation (\ref{reduit2})}
The third term is
\begin{eqnarray*}
-i \sum \big(D\i{1'}
\delta_\barx W^1_\rho\big) \big(D\i{2'} \delta_y Z_\rho\big).
\end{eqnarray*}
For this term, two kinds of $D\i{1'}\otimes D\i{2'}$
intervene: with one or three products in $D\i{1'}$.
For the first kind of terms, the relevant coproducts are
$\delta_{1}\otimes \delta_{1'}\delta_{\barunp}
\delta_{\barun}
-\delta_{1'}\otimes \delta_{1}\delta_{\barunp}
\delta_{\barun}$ and the result is
\begin{eqnarray*}
&&-\frac{i}{2}\int \dd 1 \dd 1' v(1-1') \sum_{\pm}\pm\Big(
  \big(\delta_{1}\delta_{\barx} W^1_\rho)
  \big(\delta_{1'}\delta_{\barunp}
       \delta_{\barun}\delta_{y} Z_\rho\big)
-   \big(\delta_{1'}\delta_{\barx} W^1_\rho)
  \big(\delta_{1}\delta_{\barunp}
       \delta_{\barun}\delta_{y} Z_\rho\big)\Big)
\\&&=
-i\int \dd 1 \dd 1' v(1-1') \sum_{\pm}\pm
  \big(\delta_{1}\delta_{\barx} W^1_\rho)
  \big(\delta_{1'}\delta_{\barunp}
       \delta_{\barun}\delta_{y} Z_\rho\big).
\end{eqnarray*}
We have interchanged the variables $1$ and $1'$ to obtain
the last line.
We use now 
$\delta_{1}\delta_{\barx} W^1_\rho=-iG^0_\rho(x,1)$ and
\begin{eqnarray*}
\delta_{1'}\delta_{\barunp} \delta_{\barun}\delta_{y} Z_\rho
&=& - \delta_{\barun}\delta_{\barunp}\delta_{1'}\delta_{y} Z_\rho
= G(1,1',1',y),
\end{eqnarray*}
to get
\begin{eqnarray*}
-\int \dd 1 \dd 1' v(1-1') \sum_{\pm}\pm
  G^0_\rho(x,1)G(1,1',1',y).
\end{eqnarray*}
For the second kind of terms, the relevant coproducts are
$\delta_{1}\delta_{1'}\delta_{\barunp}\otimes
\delta_{\barun}
- \delta_{1}\delta_{1'}\delta_{\barun}\otimes
   \delta_{\barunp}$. Again, interchanging $1$ and $1'$
shows that these two terms give the same contribution
and the result is
\begin{eqnarray*}
&&i\int \dd 1 \dd 1' v(1-1') \sum_{\pm} \pm
  \big(\delta_{1}\delta_{1'}\delta_{\barun}
       \delta_{\barx} W^1_\rho)
  \big(\delta_{\barunp}\delta_{y} Z_\rho\big).
\end{eqnarray*}
We calculate
\begin{eqnarray*}
\delta_{1}\delta_{1'}\delta_{\barun} \delta_{\barx} W^1_\rho
&=&  -\delta_{\barx}\delta_{\barun}\delta_{1}\delta_{1'}W^1_\rho
= \pm\epsilon r(x,1,1,1'),
\end{eqnarray*}
and
$\delta_{\barunp}\delta_{y} Z_\rho= i G(1',y)$. The contribution
of the second kind of terms becomes
\begin{eqnarray*}
-\epsilon\int \dd 1 \dd 1' v(1-1') r(x,1,1,1')\sum_{\pm}
  G(1',y).
\end{eqnarray*}

\subsection{Fourth term of equation (\ref{reduit2})}
The fourth term of equation (\ref{reduit2}) is
\begin{eqnarray*}
- \frac{1}{2}
\sum \big(D\i{1'}^{2}
\delta_y \delta_\barx W^1_\rho \big)
\big(D\i{2'}^{2} Z_\rho\big).
\end{eqnarray*}
Here we meet something new because we must use the
reduced coproduct $\Delta' D^2$. According to the
general formula (\ref{Delta'def})of I, this reduced coproduct
is obtained by multiplying $\Delta'D$ 
by itself. The spacetime variables
of the first $\Delta'D$ will still be denoted by
$1$ and $1'$, while the spacetime variables of the
second $\Delta'D$ will be denoted by
$2$ and $2'$. Moreover, the $\pm$ variable of the
second $\Delta'D$ will be written $\pm'$.
Each term $D\i{1'}$ of $\Delta'D$ is at least of degree 1,
so each term of $D\i{1'}^2$ is at least of degree 2.
Therefore, the only term of $W^1_\rho$ that contributes
to $D\i{1'}^{2} \delta_y \delta_\barx W^1_\rho$
is $\rho^c(\bara,\alpha)$. We use again the fact
that $\rho^c(\bara,\alpha)$ is a product of two
sources $\eta$ and two sources $\bareta$ to deduce that
$D\i{1'}^2$ must be the product of a functional derivative
with respect to $\eta$ and a functional derivative
with respect to $\bareta$. The eight terms $D\i{1'}^2$
that satisfy this condition are
$\delta_{1}\delta_{\barde}$,
$\delta_{1}\delta_{\bardep}$,
$\delta_{1'}\delta_{\barde}$,
$\delta_{1'}\delta_{\bardep}$,
$\delta_{1}\delta_{\barde}$,
$\delta_{\barun}\delta_{2}$,
$\delta_{\barun}\delta_{2'}$,
$\delta_{\barunp}\delta_{2}$ and
$\delta_{\barunp}\delta_{2'}$.
This gives us eight terms which are all identical if we
interchange 1 with 1' or 2 with 2' or (1,1') with (2,2').
Thus, the fourth term of equation (\ref{reduit2}) becomes
\begin{eqnarray*}
-\int \dd1\dd1'\dd2\dd2' v(1-1') v(2-2')\sum_{\pm\pm'} \pm\pm'
\big(\delta_{1}\delta_{\barde}
     \delta_{y}\delta_{\barx} W^1_\rho\big)
\big(\delta_{1'}\delta_{\barunp}\delta_{\barun}
     \delta_{2}\delta_{2'}\delta_{\bardep}
    Z_\rho\big).
\end{eqnarray*}
We have just to calculate
\begin{eqnarray*}
\delta_{1}\delta_{\barde} \delta_{y}\delta_{\barx} W^1_\rho
&=& \delta_{\barx}\delta_{\barde}\delta_{1} \delta_{y} W^1_\rho
= -\epsilon\epsilon'\pm\pm' r(x,2,1,y),
\end{eqnarray*}
and
\begin{eqnarray*}
\delta_{1}\delta_{\barunp}\delta_{\barun}
     \delta_{2}\delta_{2'}\delta_{\bardep}
    Z_\rho
&=& \delta_{\bardep}\delta_{\barun}\delta_{\barunp}
    \delta_{2}\delta_{2'}\delta_{1'}  Z_\rho
= -iG(2',1,1',2,2',1').
\end{eqnarray*}
The final expression for the fourth term is then
\begin{eqnarray*}
-i\epsilon\epsilon'\int \dd1\dd1'\dd2\dd2' v(1-1') v(2-2')
   r(x,2,1,y) \sum_{\pm\pm'} G(2',1,1',2,2',1').
\end{eqnarray*}

\subsection{Fifth term of equation (\ref{reduit2})}
The fifth term of equation  (\ref{reduit2}) is
\begin{eqnarray*}
- \frac{1}{2}
\sum \big(D\i{1'}^{2} \delta_\barx W^1_\rho\big)
\big(D\i{2'}^{2} \delta_y Z_\rho\big).
\end{eqnarray*}
To give nonzero contributions, $D\i{1'}^{2}$
must contain three functional derivatives:
one functional derivative with
respect to a $\bareta$ and two with respect to $\eta$.
This gives us three kinds of terms: (i) the 
argument of $\bareta$ does not come from the same $D$
as the arguments of the two $\eta$,
as in
$D\i{1'}^2=\delta_{\barunp}\delta_{2}\delta_{2'}$
(ii) the argument of $\bareta$ is the same as
that of one of the two $\eta$, for example
$D\i{1'}=\delta_{1}\delta_{2'}\delta_{\bardep}$
(iii) the argument of $\bareta$ comes from the
same $D$ as the argument of one of the two $\eta$
but they are not identical, for example
$D\i{1'}=-\delta_{1}\delta_{2}\delta_{\bardep}$.
In $\Delta' D^2$ there are four terms of type (i),
eight terms of type (ii)
and eight terms of type (iii).
All the terms of each class give the same contribution.
We calculate now the contribution of the three kinds of terms

\subsubsection{Terms of kind (i)}

The sum of the terms of type (i) is
\begin{eqnarray*}
-\frac{1}{2}\int \dd 1 \dd 1' \dd 2 \dd 2' v(1-1') v(2-2')
\sum_{\pm,\pm'} \pm\pm'
\big( \delta_{\barunp}\delta_{2}\delta_{2'}
      \delta_{\barx} W^1_\rho\big)
\big( \delta_{1}\delta_{1'}
      \delta_{\barun}
      \delta_{\bardep}\delta_{\barde}
      \delta_{y} Z_\rho\big).
\end{eqnarray*}
We compute
\begin{eqnarray*}
\delta_{\barunp}\delta_{2}\delta_{2'}
      \delta_{\barx} W^1_\rho
&=& - \delta_{\barx}\delta_{\barunp}\delta_{2}
         \delta_{2'} W^1_\rho
= \pm\epsilon r(x,1',2,2'),
\end{eqnarray*}
and
\begin{eqnarray*}
\delta_{1}\delta_{1'}
      \delta_{\barun}
      \delta_{\bardep}\delta_{\barde}
      \delta_{y} Z_\rho
&=& - \delta_{\barun}\delta_{\barde}\delta_{\bardep}
     \delta_{1}\delta_{1'}\delta_{y} Z_\rho
= i G(1,2,2',1,1',y)
\end{eqnarray*}
to obtain
\begin{eqnarray*}
-\frac{i}{2}\epsilon\int \dd 1 \dd 1' \dd 2 \dd 2' v(1-1') v(2-2')
  r(x,1',2,2') \sum_{\pm,\pm'} \pm'  G(1,2,2',1,1',y).
\end{eqnarray*}

\subsubsection{Terms of kind (ii)}

The sum of the terms of type (ii) is
\begin{eqnarray*}
-\int \dd 1 \dd 1' \dd 2 \dd 2' v(1-1') v(2-2')
\sum_{\pm,\pm'} \pm\pm'
\big(\delta_{1}\delta_{2'}\delta_{\bardep}
      \delta_{\barx} W^1_\rho\big)
\big( \delta_{1'}\delta_{\barunp}
      \delta_{\barun}
      \delta_{2}\delta_{\barde}
      \delta_{y} Z_\rho\big).
\end{eqnarray*}
We use
\begin{eqnarray*}
\delta_{1}\delta_{2'}\delta_{\bardep}
      \delta_{\barx} W^1_\rho
&=& - \delta_{\barx}\delta_{\bardep}\delta_{1}\delta_{2'}W^1_\rho
=\pm\epsilon r(x,2',1,2'),
\end{eqnarray*}
and
\begin{eqnarray*}
\delta_{1'}\delta_{\barunp}
      \delta_{\barun}
      \delta_{2}\delta_{\barde}
      \delta_{y} Z_\rho
&=&
-\delta_{\barun}\delta_{\barunp}\delta_{\barde}
 \delta_{1'}      \delta_{2}\delta_{y} Z_\rho
= i G(1,1',2,1',2,y).
\end{eqnarray*}
Therefore, the fifth term of type (ii) becomes
\begin{eqnarray*}
-i\epsilon\int \dd 1 \dd 1' \dd 2 \dd 2' v(1-1') v(2-2')
  r(x,2',1,2')
\sum_{\pm,\pm'} \pm' G(1,1',2,1',2,y).
\end{eqnarray*}

\subsubsection{Terms of kind (iii)}

The sum of the terms of type (iii) is
\begin{eqnarray*}
\int \dd 1 \dd 1' \dd 2 \dd 2' v(1-1') v(2-2')
\sum_{\pm,\pm'}\pm\pm'
\big(\delta_{1}\delta_{2}\delta_{\bardep}
      \delta_{\barx} W^1_\rho\big)
\big( \delta_{1'}\delta_{\barunp}
      \delta_{\barun}
      \delta_{2'}\delta_{\barde}
      \delta_{y} Z_\rho\big).
\end{eqnarray*}
We have
\begin{eqnarray*}
\delta_{1}\delta_{2}\delta_{\bardep}
      \delta_{\barx} W^1_\rho
&=&\delta_{\barx}\delta_{\bardep}\delta_{1}\delta_{2}W^1_\rho
=\mp\epsilon r(x,2',1,2),
\end{eqnarray*}
and
\begin{eqnarray*}
\delta_{1'}\delta_{\barunp}
      \delta_{\barun}
      \delta_{2'}\delta_{\barde}
      \delta_{y} Z_\rho
&=&
- \delta_{\barun}\delta_{\barunp}\delta_{\barde}
\delta_{1'}      \delta_{2'}      \delta_{y} Z_\rho
= i G(1,1',2,1',2',y).
\end{eqnarray*}
The fifth term of type (iii) becomes
\begin{eqnarray*}
-i\epsilon \int \dd 1 \dd 1' \dd 2 \dd 2' v(1-1') v(2-2')
    r(x,2',1,2)
\sum_{\pm,\pm'}\pm' G(1,1',2,1',2',y).
\end{eqnarray*}

\subsection{Sixth term of equation (\ref{reduit2})}
The sixth term of equation (\ref{reduit2}) is
\begin{eqnarray*}
 \frac{i}{6}
\sum \big(D\i{1'}^{3} \delta_\barx W^1_\rho\big)
\big(D\i{2'}^{3} \delta_y Z_\rho \Big).
\end{eqnarray*}
It looks complicated because we have now
$\Delta' D^3=(\Delta'D)(\Delta'D)(\Delta'D)$.
The third $\Delta'D$ will have the 
variables $3$, $3'$ and $\pm''$.
Again, we need one derivative with respect to
$\bareta$ and two derivatives with respect to $\eta$
in $D\i{1'}^{3}$. There are 24 terms of this kind
in $\Delta' D^3$, and they all give the same contribution
as 
$\delta_\barun\delta_2\delta_3\otimes
  \delta_1\delta_{1'}\delta_{\barunp}\delta_{2'}\delta_{\bardep}
    \delta_{\barde} \delta_{3'}\delta_{\bartrp} \delta_{\bartr}$.
Thus, the sixth term is
\begin{eqnarray*}
&&\frac{i}{2}\int \dd 1 \dd 1' \dd 2 \dd 2' \dd 3 \dd 3'
    v(1-1') v(2-2') v(3-3') \sum_{\pm\pm'\pm''}\pm\pm'\pm''
  \big(\delta_\barun\delta_2\delta_3\delta_\barx W^1_\rho\big)
  \big(\delta_1\delta_{1'}\delta_{\barunp}\delta_{2'}\delta_{\bardep}
    \delta_{\barde} \delta_{3'}\delta_{\bartrp} \delta_{\bartr}
    \delta_y Z_\rho\big).
\end{eqnarray*}
We compute
\begin{eqnarray*}
\delta_\barun\delta_2\delta_3\delta_\barx W^1_\rho
&=& - \delta_\barx\delta_\barun\delta_2\delta_3 W^1_\rho
= \pm\pm'\pm''\epsilon r(x,1,2,3),
\end{eqnarray*}
and
\begin{eqnarray*}
\delta_1\delta_{1'}\delta_{\barunp}\delta_{2'}\delta_{\bardep}
    \delta_{\barde} \delta_{3'}\delta_{\bartrp} \delta_{\bartr}
    \delta_y Z_\rho
&=& 
\delta_{\barunp}\delta_{\barde}\delta_{\bardep}\delta_{\bartr}
   \delta_{\bartrp}\delta_1\delta_{1'}\delta_{2'}\delta_{3'}
   \delta_y Z_\rho
=
 i G(1',2,2',3,3',1,1',2',3',y).
\end{eqnarray*}
The last term of equation (\ref{reduit2}) is then
\begin{eqnarray*}
&&-\frac{1}{2}\epsilon\int \dd 1 \dd 1' \dd 2 \dd 2' \dd 3 \dd 3'
    v(1-1') v(2-2') v(3-3')  r(x,1,2,3)
  \sum_{\pm\pm'\pm''} G(1',2,2',3,3',1,1',2',3',y).
\end{eqnarray*}

\subsection{The Green function hierarchy}
If we gather all the previous results, we obtain
the Green function hierarchy
\begin{eqnarray}
G(x,y) &=& G_{\epsilon\epsilon'}(x,y) =
G^0_{\rho\epsilon\epsilon'}(x,y) 
-i\int \dd 1 \dd 1' v(1-1') \sum_{\pm}\pm
  G^0_\rho(x,1)G(1,1',1',y)
\nonumber\\&&
-i\epsilon\epsilon' \int \dd 1 \dd 1' v(1-1') r(x,1',1,y)
\sum_{\pm} \pm G(1,1')
+ i\epsilon\epsilon' \int \dd 1 \dd 1' v(1-1') r(x,1,1,y)
\sum_{\pm} \pm G(1',1')
\nonumber\\&&
-i\epsilon\int \dd 1 \dd 1' v(1-1') r(x,1,1,1')\sum_{\pm}
  G(1',y)
\nonumber\\&&
+\frac{1}{2}\epsilon\int \dd 1 \dd 1' \dd 2 \dd 2' v(1-1') v(2-2')
  r(x,1',2,2') \sum_{\pm,\pm'} \pm'  G(1,2,2',1,1',y)
\nonumber\\&&
+\epsilon \int \dd 1 \dd 1' \dd 2 \dd 2' v(1-1') v(2-2')
    r(x,2',1,2)
\sum_{\pm,\pm'}\pm' G(1,1',2,1',2',y).
\nonumber\\&&
+\epsilon\epsilon'\int \dd1\dd1'\dd2\dd2' v(1-1') v(2-2')
   r(x,2,1,y) \sum_{\pm\pm'} G(2',1,1',2,2',1')
\nonumber\\&&
+\epsilon\int \dd 1 \dd 1' \dd 2 \dd 2' v(1-1') v(2-2')
  r(x,2',1,2')
\sum_{\pm,\pm'} \pm' G(1,1',2,1',2,y).
\nonumber\\&&
-\frac{i}{2}\epsilon\int \dd 1 \dd 1' \dd 2 \dd 2' \dd 3 \dd 3'
    v(1-1') v(2-2') v(3-3')  r(x,1,2,3)
   \sum_{\pm\pm'\pm''} G(1',2,2',3,3',1,1',2',3',y).
\label{hierarchie}
\end{eqnarray}

In terms of the Feynman diagrams of nonequilibrium
quantum field theory \cite{Hall}, these terms can
be rewritten as in figure 1. The order of the terms is
the same in equation (\ref{hierarchie}) and in figure 1.
\begin{figure}
\includegraphics*{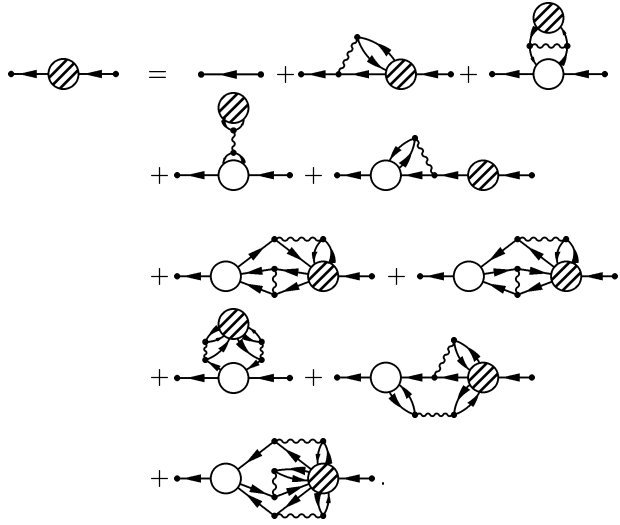}
\caption{Diagrammatic representation of equation (\ref{hierarchie}).}
\end{figure}

\section{Conclusion}

The hierarchy of Green functions presented here is 
an important step in the effective calculation
of degenerate systems. However, the practical
implementation of this hierarchy requires an
approximation to close it. The GW 
approximation \cite{Hedin,GW,Arya}
is a well-known and powerful way of closing
the hierarchy. It has to be adapted to
our more complicated hierarchy.
For the calculation of the optical properties
of allochromatic crystals, it will also be
necessary to adapt the Bethe-Salpeter
approach \cite{Onida} to our setting.

Moreover, a functional derivation of the energy
with respect to the density matrix provides
equations that enable us to unify the Green-function
formalism and the diagonalization method of
many-body theory. This will be presented in
a forthcoming publication \cite{BrouderX}.

\end{document}